# A New Statistic Feature of the Short-Time Amplitude Spectrum Values for Human's Unvoiced Pronunciation

XIAODONG ZHUANG[1]

1. Qingdao University, Electronics & Information College, Qingdao, 266071 CHINA

*Abstract:* - In this paper, a new statistic feature of the discrete short-time amplitude spectrum is discovered by experiments for the signals of unvoiced pronunciation. For the random-varying short-time spectrum, this feature reveals the relationship between the amplitude's average and its standard for every frequency component. On the other hand, the association between the amplitude distributions for different frequency components is also studied. A new model representing such association is inspired by the normalized histogram of amplitude. By mathematical analysis, the new statistic feature discovered is proved to be necessary evidence which supports the proposed model, and also can be direct evidence for the widely used hypothesis of "identical distribution of amplitude for all frequencies".

## 1 Introduction

Speech signal can be mathematically modelled by stochastic process. The speech features are random and time-varying in both time domain and transformed domains such as the short-time spectrum [1,2]. The statistic feature of speech signal is one of the important research topics. In the frequency domain, the short-time amplitude spectrum values can be mathematically taken as random variables, and there have been researches estimating their probability distribution, which facilitates the application of speech enhancement [3,4]. Such researches are based on the large amount of speech data in corpora like TIMIT or other database of daily speech signal from the internet [2,5].

However, these studies are based on the words or sentences spoken in daily-life communication, which are the mixture of various pronunciation types including vowel, consonant, plosive, etc. Based on such corpora, the estimated statistic feature is in fact the overall feature of the signal mixed by different pronunciation types. Therefore, it is necessary to further study the statistic feature of specific pronunciation type (or specific phoneme) alone, because different types have different pronunciation mechanisms.

The unvoiced pronunciation is one of the major pronunciation types, which is closely related to the aerodynamic process in vocal tract [6-8]. The physical process of unvoiced pronunciation is complicated, while the statistical study of its signal may reveal some underlying properties of it. In this paper, the statistic study is carried out in the frequency domain for unvoiced pronunciation. A novel statistical feature named "consistent standard deviation coefficient" is discovered for short-time amplitude spectrum data, which is revealed by the statistic study on stable and sustained signals of unvoiced pronunciation. Moreover, the relationship between the amplitude probability distributions of two different frequency components is investigated, based on which a new model is proposed representing such relationship. The validity of the new model is supported in mathematical analysis with the discovered statistic feature as direct evidence, which has potential application like speech synthesis.

## 2 New Statistic Feature in Frequency Domain for Unvoiced Pronunciation

In order to obtain sufficient data for statistic study, the signals used in this study are stable and sustained pronunciations. For each unvoiced phoneme studied, its signal is recorded, and each signal is studied alone. For each signal, the short-time Fourier transform (STFT) is used to gather sufficient spectrum data for the statistic study. Since the STFT used is in discrete form, the spectrum has finite number of discrete components, and the statistic study is eventually performed for each frequency component individually.

Since currently there is little corpus of sustained phoneme pronunciation, signals have been captured using microphones connected to the sound card on computers. The signals were recorded at sample frequency of 16 kHz, with 16 bit per sample. To guarantee the generality of experimental results, signals have been captured for a group of unvoiced pronunciation spoken by different speakers, and on different recording platforms (different microphones and sound cards on different computers). In the collection of signal, the speakers were informed with the requirements of stable pronunciation during sufficient time length, which is required by reliable statistic study. For each unvoiced phoneme, the stability of pronunciation largely determines the effectiveness of further analysis, therefore the signals were captured repeatedly for several times, and the most stable signal can be selected.

In the STFT on each signal, the frame length is set to 512, which corresponds to a time interval of 32ms for a 16 kHz sampling frequency. A Hamming window is used on each frame in STFT. Let $\omega_k$ denotes the $k$-th frequency component in STFT. Due to the randomness of the signal, the amplitude of $\omega_k$ also varies randomly in each frame of the signal. Let $\mu(\omega_k)$ and $\sigma^2(\omega_k)$ represent the estimated average and variance of $\omega_k$'s amplitude respectively. And the estimated standard deviation $\sigma(\omega_k)$ is the square-root of $\sigma^2(\omega_k)$. Mathematically, $\mu(\omega_k)$ and $\sigma(\omega_k)$ are two functions, and their curves can be drawn after $\mu$ and $\sigma$ are estimated for each frequency $\omega_k$. For a dozen of unvoiced phoneme, the above basic statistic is estimated. Some typical results are shown in Fig. 1 and Fig. 2 as the curves of $\mu(\omega_k)$ and $\sigma(\omega_k)$. It can be observed evidently that there is clear similarity between the curves of $\mu(\omega_k)$ and $\sigma(\omega_k)$. Such similarity also exists in all the other results of unvoiced pronunciation in the experiments, which inspires the study of the relationship between the two function $\mu(\omega_k)$ and $\sigma(\omega_k)$ as following.

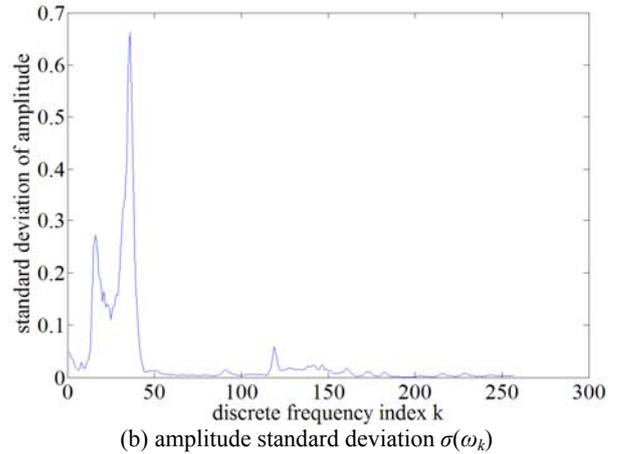

(b) amplitude standard deviation $\sigma(\omega_k)$

Fig. 1. The estimated expectation and standard deviation of the short-time amplitude spectrum for [h]

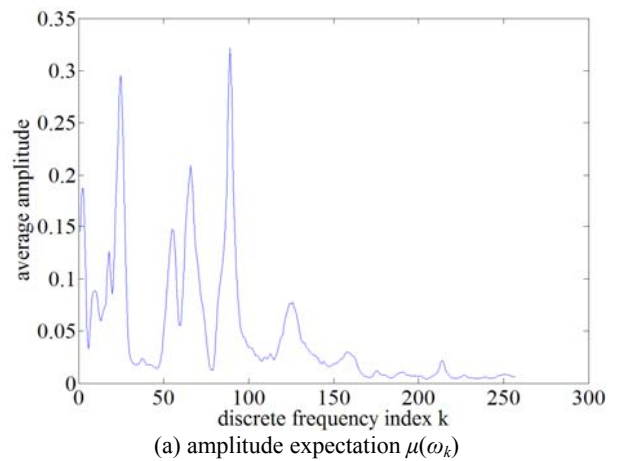

(a) amplitude expectation $\mu(\omega_k)$

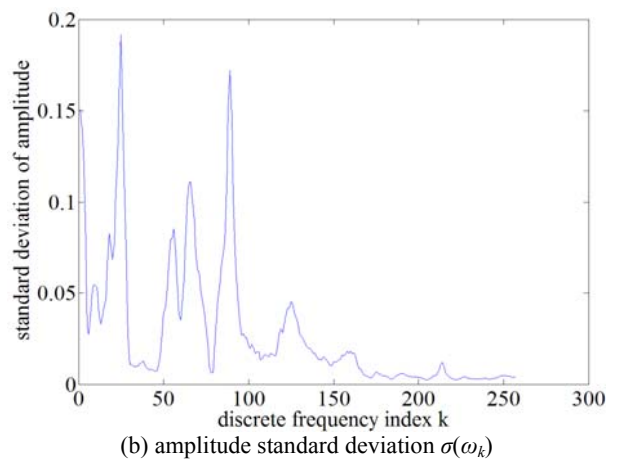

(b) amplitude standard deviation $\sigma(\omega_k)$

Fig. 2. The estimated expectation and standard deviation of the short-time amplitude spectrum for unvoiced [e]

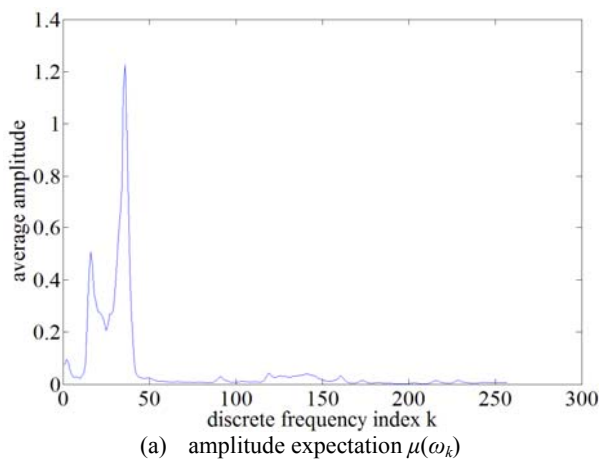

(a) amplitude expectation $\mu(\omega_k)$

Besides the above experimental results, the relationship between $\mu(\omega_k)$ and $\sigma(\omega_k)$ is quantitatively verified by calculating the correlation coefficient between the two curves of $\mu(\omega_k)$ and $\sigma(\omega_k)$. The correlation coefficient is calculated in a discrete form:

$$\rho_{\sigma\mu} = \frac{\sum_{k=1}^{N} \sigma(\omega_k) \cdot \mu(\omega_k)}{\sqrt{\sum_{k=1}^{N} \sigma^2(\omega_k)} \cdot \sqrt{\sum_{k=1}^{N} \mu^2(\omega_k)}} \qquad (1)$$

where $N$ is the number of discrete frequencies in the discrete spectrum. Some of the experimental results are shown in Table 1, which are based on the pronunciation signals recorded for one male speaker. The correlation coefficients between $\mu(\omega_k)$ and $\sigma(\omega_k)$ are calculated for different unvoiced phonemes. The correlation coefficients between $\mu(\omega_k)$ and $\sigma(\omega_k)$ are much close to 1.0. Consider the unavoidable error caused by the instability of sustained natural pronunciation, and also the noise introduced in the signal capture process, the results indicate that $\mu(\omega_k)$ and $\sigma(\omega_k)$ are strongly related by a linear proportional relationship, which is a new statistic feature discovered for human's unvoiced pronunciation.

Table 1  The correlation coefficient of $\mu(\omega_k)$ and $\sigma(\omega_k)$ for unvoiced pronunciation

| Pronunciation | $\rho$ between $\mu(\omega_k)$ and $\sigma(\omega_k)$ | Number of signal frames |
|---|---|---|
| [s] (male) | 0.9910 | 35748 |
| [θ] (male) | 0.9852 | 28126 |
| [f] (male) | 0.9948 | 40179 |
| [h] (male) | 0.9982 | 21909 |
| unvoiced [a] (male) | 0.9960 | 17497 |
| unvoiced [ə] (male) | 0.9817 | 45336 |
| unvoiced [e] (male) | 0.9913 | 41872 |
| unvoiced [i] (male) | 0.9896 | 44147 |

Because the parameter of "the standard deviation coefficient" represents the $\sigma$ to $\mu$ ratio, the above statistic feature is named as the feature of "consistent standard deviation coefficient". In another word, for the pronunciation of an unvoiced phoneme, the proportional coefficient between the standard deviation and the expectation is consistent for all the frequency components in the short-time amplitude spectrum. This feature can also be expressed by:

$$\sigma(\omega_k) = c_s \cdot \mu(\omega_k) \qquad (2)$$

where $c_s$ is the consistent standard deviation coefficient of amplitude for all frequency components. The subscript $s$ means that Equation (2) is for one signal of unvoiced pronunciation. If the signal is changed to the one of another different unvoiced pronunciation, the value $c_s$ may also change.

Because the expectation and the standard deviation are two basic statistic of a random variable, the feature of "consistent standard deviation coefficient" indicates that there is certain association between the amplitude probability distributions of different frequency components, which is studied in the next section.

## 3 The Relationship between Amplitude Probability Distributions of Different Frequency Components

Based on the spectrum data obtained by STFT, the histogram of amplitude for each frequency component $\omega_k$ is computed. The histogram reflects the distribution of random amplitude data for each $\omega_k$, which is closely related to the amplitude probability distribution. Therefore, the amplitude histogram of each $\omega_k$ is compared to those of other frequencies, in order to study the relationship between the corresponding probability distributions.

On the other hand, in order to study the amplitude distribution type of different $\omega_k$ without the influence of different average value, the normalized histogram is also computed for each $\omega_k$. The normalization is for the average of amplitude. First, the average of amplitude for $\omega_k$ is computed. After that, each amplitude data of $\omega_k$ is divided by that average value as a preprocessing step. The normalized histogram is then computed based on the data after that preprocessing.

For a dozen of unvoiced phonemes, the original histogram and normalized histogram of amplitude are both computed for comparison. Two typical results are shown in Fig. 3 and Fig. 4. In order to find clues of the relationship between amplitude distributions of different $\omega_k$, the histogram curves of every $\omega_k$ are plotted together as a family of curves.

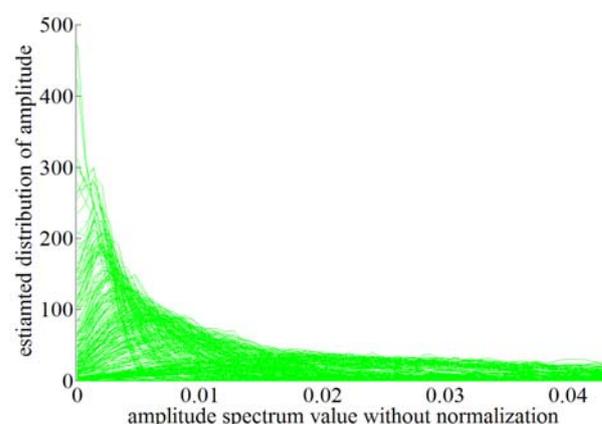

(a) The amplitude histograms before amplitude normalization

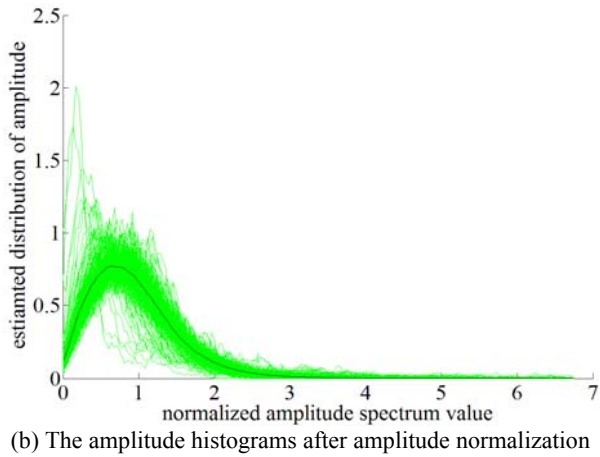

(b) The amplitude histograms after amplitude normalization

Fig. 3. The amplitude histogram of each frequency $\omega_k$ for [h]

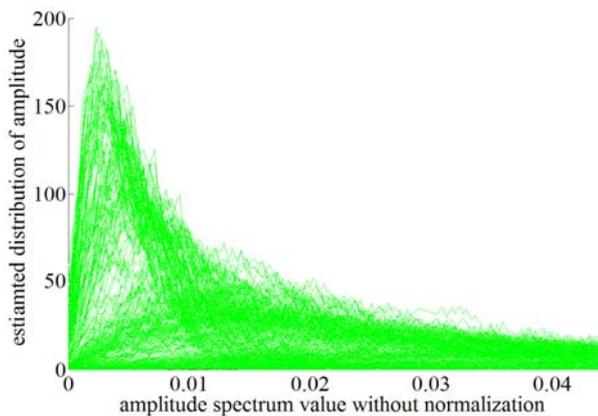

(a) The amplitude histograms before amplitude normalization

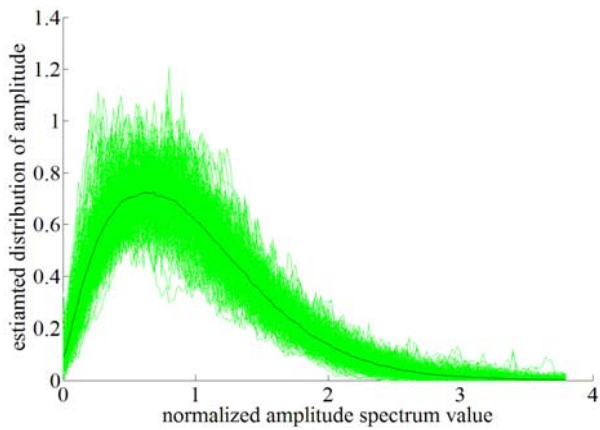

(b) The amplitude histograms after amplitude normalization

Fig. 4. The amplitude histogram of each frequency $\omega_k$ for unvoiced [e]

In Fig. 3(a) and Fig. 4(a), the original histogram curves are mixed and there is no obvious regularity between them. However, in Fig. 3(b) and Fig. 4(b), the normalized histogram curves obviously converge to one central curve (shown in black colour), especially compared to (a) of these figures. Because the normalized histogram curves converge closely, the mixed plotting results in a belt around a central curve. For other unvoiced phonemes, similar results are obtained. The results indicate the strong association between the amplitude distributions of different $\omega_k$.

Based on the above results, a new model of amplitude distribution in frequency domain is proposed for human's unvoiced pronunciation. In the model, for the signal of some unvoiced pronunciation, the amplitude distributions for different $\omega_k$ are of the same type, but with different expectation (or average) values. In another word, there is a prototype distribution function $p_0(a_0)$, from which the amplitude distribution of any $\omega_k$ can be derived by varying the expectation. The prototype $p_0(a_0)$ corresponds to the central curve (in black colour) in Fig. 3(b) or Fig. 4(b). This model can also be described mathematically as follows. As a random variable $a$, the amplitude of some $\omega_k$ is modeled as the scaling of a prototype random variable $a_0$, whose expectation is 1:

$$a = k \cdot a_0 \quad (3)$$

where $k$ is the scaling parameter. Equation (3) is a mathematical description of the model proposed. In the model, $a_0$ is the same for each frequency component, but the scaling parameter $k$ may be different for different $\omega_k$.

Besides the normalized amplitude histograms as direct inspiration of the model, it can also find proof from the new discovered statistic feature in Section 2. In the following, the feature of "consistent standard deviation coefficient" can be theoretically induced from the proposed model; in another word, this model accords well with the feature of "consistent standard deviation coefficient" discovered in the experiments. First, consider the probability distribution of $a$ in Equation (3), given $p_0(a_0)$ is the probability distribution of $a_0$. According to Equation (3), the expectation of $a$ is:

$$\mu_a = E[a] = E[k \cdot a_0] = k \cdot E[a_0] = k \cdot \mu_0 \quad (4)$$

where $\mu_0$ is the expectation of $a_0$. Based on the pdf (probability distribution function) of a variable's function in probability theory, the probability distribution of $a$ can be deduced as:

$$p(a) = \frac{1}{k} \cdot p_0\left(\frac{a}{k}\right) \quad (5)$$

Second, consider the standard deviation coefficient of $a$:

$$\frac{\sigma_a}{\mu_a} = \frac{\sqrt{Var(a)}}{\mu_a} = \frac{\sqrt{\int_{-\infty}^{+\infty}(a-\mu_a)^2 p(a)da}}{\mu_a} \quad (6)$$

Considering Equation (4) and (5), Equation (6) can be rewritten as:

$$\frac{\sigma_a}{\mu_a} = \frac{\sqrt{\int_{-\infty}^{+\infty}(a - k\cdot\mu_0)^2 \cdot \frac{1}{k} p_0\left(\frac{a}{k}\right) da}}{k\cdot\mu_0} \quad (7)$$

Then do the variable substitution $a=ka_0$ to the integral on the right side of Equation (7):

$$\frac{\sigma_a}{\mu_a} = \frac{\sqrt{\int_{-\infty}^{+\infty}(ka_0 - k\mu_0)^2 \cdot \frac{1}{k} p_0(a_0) d(ka_0)}}{k\mu_0}$$

$$= \frac{\sqrt{k^2} \cdot \sqrt{\int_{-\infty}^{+\infty}(a_0 - \mu_0)^2 \cdot p_0(a_0) da_0}}{k\mu_0} \quad (8)$$

Remember that the variables $a$ and $a_0$ represent the amplitude value, which is non-negative. Therefore, $k$ is also non-negative. Then Equation (8) can be rewritten as:

$$\frac{\sigma_a}{\mu_a} = \frac{\sqrt{\int_{-\infty}^{+\infty}(a_0 - \mu_0)^2 \cdot p_0(a_0) da_0}}{\mu_0} \quad (9)$$

Notice that the numerator of the right side of Equation (9) is just the standard deviation of $a_0$. Therefore,

$$\frac{\sigma_a}{\mu_a} = \frac{\sigma_0}{\mu_0} \quad (10)$$

Notice that the right side of Equation (10) is constant given the prototype distribution $p_0(a_0)$. Therefore, the standard deviation coefficient of $a$ is consistent whatever the scaling factor $k$ is, which is equal to that of the prototype variable $a_0$. This just accords well with the experimental results shown in Section 2. Therefore, the feature of "consistent standard deviation coefficient" supports the model proposed here.

## 4 Conclusion

In this paper, the statistic feature of unvoiced pronunciation in frequency domain is studied. The Study is focused on the short-time amplitude spectrum, and is based on the data obtained by STFT on signals of stable and sustaining unvoiced pronunciations. A new statistic feature named "consistent standard deviation coefficient" is discovered. This feature indicates strong associations between amplitude distributions of different frequency components. On the other hand, such association is also revealed by comparing the normalized amplitude histograms of every frequency components. A new model is proposed to representing such association. In this model, the random variables representing amplitude of every frequency component belong to the same pdf type, but they have different expectations. If the prototype pdf $p_0(a_0)$ is determined, the pdf of any frequency's amplitude $a$ can be derived by $a=\mu a_0$, where $\mu$ is $a$'s expectation. Moreover, by mathematical analysis, this model accords well with the feature of "consistent standard deviation coefficient". The results in the paper deepen the understanding of the stochastic features of unvoiced pronunciation, which is an important topic in speech signal analysis. In future work, the specific pdf type will be studied to suit the short-time amplitude spectrum data for unvoiced pronunciation. And other types of pronunciation like voiced phonemes will be also studied statistically for new possible features.